\documentstyle[sprocl,epsf]{article}

\def\Journal#1#2#3#4{{#1} {\bf #2}, #3 (#4)}

\def\NPBS{{\em Nucl. Phys.} B (Proc.\ Suppl.)}
\def\PLB{{\em Phys. Lett.}  B}

\def\PRD{{\em Phys. Rev.} D}
\def\AP{{\em Ann. Phys.}}
\def\ZPC{{\em Z. Phys.} C}

\def\ra{\rightarrow}

\def\be{\begin{equation}}
\def\ee{\end{equation}}
\def\bea{\begin{eqnarray}}
\def\eea{\end{eqnarray}}

\def\beeq{\begin{equation}}
\def\eneq{\end{equation}}
\def\beqn{\begin{eqnarray}}
\def\eeqn{\end{eqnarray}}

\def\mybig{\displaystyle \strut }

\def\dd{\partial}
\def\la{\raise.16ex\hbox{$\langle$}\lower.16ex\hbox{}  }
\def\ra{\, \raise.16ex\hbox{$\rangle$}\lower.16ex\hbox{} }
\def\go{\rightarrow}

\def\onehalf{ \hbox{${1\over 2}$} }

\def\psibar{ \psi \kern-.65em\raise.6em\hbox{$-$} }
\def\chibar{ \chi \kern-.65em\raise.5em\hbox{$-$} }
\def\mbar{ m \kern-.75em\raise.4em\hbox{$-$}\hbox{} }
\def\Bbar{ B \kern-.73em\raise.6em\hbox{$-$}\hbox{} }

\def\ep{\epsilon}

\def\wil{ \Theta_{\rm W} }

\def\vphi{ {\varphi} }

\def\eff{{\rm eff}}

\def\Lap{{\triangle}}
\def\LapN{{\triangle_N}}
\def\potN{{V_N}}

\def\boxit#1{$\vcenter{\hrule\hbox{\vrule\kern3pt
     \vbox{\kern3pt\hbox{#1}\kern3pt}\kern3pt\vrule}\hrule}$}
\def\bigbox#1{$\vcenter{\hrule\hbox{\vrule\kern5pt
     \vbox{\kern5pt\hbox{#1}\kern5pt}\kern5pt\vrule}\hrule}$}
\def\hugebox#1{$\vcenter{\hrule\hbox{\vrule\kern8pt
     \vbox{\kern8pt\hbox{#1}\kern8pt}\kern8pt\vrule}\hrule}$}

\begin{document}

{\small \noindent March 21, 1997  \hfill UMN-TH-1531/97}


\vskip .5cm 

\title{CHIRAL DYNAMICS IN WEAK, INTERMEDIATE, AND STRONG COUPLING
QED IN TWO DIMENSIONS \footnote{~ To appear in the Proceedings of 
{\it 1996
International Workshop on Perspectives of Strong Coupling Gauge
Theories}, Nagoya, November 13 - 16, 1996}}

\author{Y. HOSOTANI}

\address{School of Physics and Astronomy, University of Minnesota\\
Minneapolis, MN 55455, USA}


\maketitle\abstracts{$N$ flavor QED in two dimensions is reduced to a
quantum mechanics problem with $N$ degrees of freedom for which the potential
is determined by the ground state of the problem itself.  The chiral
condensate is determined at all values of temperature, fermion masses,
and the $\theta$ parameter.  In the single flavor case, the anomalous
mass ($m$) dependence of the chiral condensate at $\theta=\pi$ at low
temperature is found.  The critical value is given by
$m_c \sim .437 \cdot e/\sqrt{\pi} $.}

\section{QED$_2$ and Quantum Mechanics}

Two-dimensional QED with massive fermions is not exactly solvable.
The relevant parameter measuring the strength of the interaction is
$e/m$ where $e$ and $m$ are the gauge coupling and fermion mass,
respectively.  The massless limit corresponds to the strong coupling,
whereas the large fermion mass limit corresponds to the weak coupling.
In both limits the model is exactly solvable.  For the intermediate
coupling $e/m={\rm O}(1)$, however, the model is highly interacting
and some approximation methods need to be employed.

The model has many similarities to the four-dimensional QCD in such respects
as  chiral dynamics and confinement.  In this article we would like
to  focus on the chiral condensate at all values of the coupling,  temperature,
and  $\theta$ parameter. 

The Lagrangian is 
\beeq
{\cal L} = - \hbox{$1\over 4$} \, F_{\mu\nu} F^{\mu\nu} + 
\sum_{a=1}^N \psibar_a \Big\{ \gamma^\mu (i \dd_\mu - e A_\mu) -
  m_a  \Big\} \psi_a ~~.
\label{Lagrangian}
\eneq
We examine the model defined on a circle $S^1$ with a circumference
$L$.  Upon imposing periodic and anti-periodic boundary conditions on
the bosonic and fermionic fields, respectively, the model is
mathematically equivalent to a theory defined on a line ($R^1$) at
finite temperature ($T$) by analytic continuation to imaginary time
$\tau$ $(=it$) and the interchange of $\tau$ and $x$.  Various physical
quantities at $T\not=0$ on $R^1$ are obtained from the corresponding
ones at $T=0$ on $S^1$ by substituting $T^{-1}$ for $L$.

In a series of papers it has been shown that the field theory system
(\ref{Lagrangian}) is effectively reduced to a quantum mechanical system 
of $N$ degrees of freedom.$^{1-5}$
The argument proceeds as follows.  

Fermions are bosonized on a circle.      Each
two-component fermion ($\psi_a$) is expressed in terms of zero modes
($q_a^\pm , p_a^\pm$) and oscillatory modes ($\phi_a(x), \Pi_a(x)$).
The boundary conditions enforce that $p_a^\pm$'s take integer
eigenvalues. 

The relevant parts of the Hamiltonian are expressed in terms of
($q_a$=$q_a^++q_a^-, p_a$=$\onehalf[p_a^++p_a^-]$), ($\phi_a,\Pi_a$),
and ($\wil, p_W$) where $\wil$ is the Wilson line phase, the only
physical degree of freedom associated with gauge fields on a circle.
In the massless fermion theory the zero modes and oscillatory modes
decouple.  There appear one massive oscillatory mode, $\chi_1$ with 
$\mu_1^2 = Ne^2/\pi$, and $N-1$ massless
oscillatory modes, $\chi_\alpha$ with $\mu_\alpha^2=0$
($\alpha=2 \sim N$).  

Fermion masses introduce nontrivial interactions among
these variables.  All of the oscillatory modes become massive.  
Fermion masses  change  the vacuum structure.  
The chiral condensate is induced and the boson mass spectrum is 
modified.  Each affects the other, and must be determined self-consistently.

There are $N+1$ relevant zero modes, $\wil$ and $q_a$.  The $\theta$ vacuum
structure, which follows from the gauge invariance,  eliminates
one degree.  With an appropriate choice of the basis, we have $N$ zero
mode degrees which we denote by $p_W$ and $\vphi_a$ ($a=1\sim N-1$).  The
vacuum wave function must solve the Schr\"odinger equation
\beqn
&&H \, f(p_W,\vphi) = \ep \, f(p_W, \vphi) \cr
\noalign{\kern 10pt}
&& H =  -{\dd^2\over \dd
p_W^2}  - {4\pi^2 (N-1)\over N^2}~ \Lap_\vphi + V_N(p_W,\vphi) 
\label{Schro1}
\eeqn
where
\beqn
\LapN &=&  \sum_{a=1}^{N-1} {\dd^2\over \dd\vphi_a^2} 
-{2\over N-1} \sum_{a<b}^{N-1} {\dd^2\over \dd\vphi_a\dd\vphi_b} \cr
\potN ~ &=& \Big( {\pi\mu L p_W\over N} \Big)^2 
- {4\pi \over N} \sum_{a=1}^N m_a L \Bbar_a 
       \cos \Big( \vphi_a - {2\pi p_W\over N}\Big) \cr
\vphi_N &=& \theta_\eff - \sum_{a=1}^{N-1} \vphi_a  ~.
\label{Schro2}
\eeqn

In the potential $\potN$, $\Bbar_a$ depends on the boson mass
spectrum $\mu_\alpha$ as well as $e$, $L$, and $m_a$.  The detailed form
has been given in refs.\ 1 and 3.   The spectrum $\mu_\alpha$ is to be
determined from the vacuum wave function solving (\ref{Schro1}). 
We have a routine
\beeq
V(p_W,\vphi) \go f(p_W, \vphi) \go \mu_\alpha^2 \go V(p_W, \vphi)~.
\label{routine}
\eneq
This is a rather non-trivial condition to be satisfied.  The Hamiltonian 
is determined by its ground state wave function. 

\section{Chiral condensates --- $T$, $m$, and $\theta$-dependence}

In the strong ($e/m \gg 1$) and weak ($e/m \ll 1$) coupling limits
Eq.\ (\ref{Schro1}) and the condition (\ref{routine}) can be solved
analytically.  For general values of the parameters, however,
they must be solved numerically.  We have numerically solved the Schr\"odinger 
equation on workstations.

Let us focus on the $N=1$ (single flavor) model.  In this case there is 
no $\vphi$ degree,  the
Sch\"odinger equation (\ref{Schro1}) becoming
\beqn
&&\bigg\{  - {d^2 \over dp_W^2} + V(p_W) \bigg\} \, f(p_W) 
 = \ep \, f(p_W)\cr
\noalign{\kern 10pt}
&& \hskip 1cm V(p_W)  =   (\pi \mu L p_W)^2 
 - \kappa \,  \cos(\theta - 2\pi p_W) \cr
\noalign{\kern 8pt}
&& \hskip 1cm \kappa  = 4\pi mLB(\mu_1 L)  \cr
\noalign{\kern 8pt}
&& \hskip 1cm B(z) = {z\over 4\pi} \, \exp \bigg( \gamma + {\pi\over z}
- 2 \int_1^\infty {du\over (e^{uz} - 1) \sqrt{u^2-1}} \bigg)
\label{Schro3}
\eeqn
where $\mu^2 = e^2/\pi$.   Note that $B(0)=1$.
$\mu_1$ is the boson mass which is determined
by the vacuum wave function $f(p_W)$:
\beeq
\mu_1^2 =  \mu^2 + {8\pi m B(\mu_1L)\over L} 
 ~  \la \cos (2 \pi p_W - \theta ) \ra_f ~.
\label{mass1}
\eneq
In the weak coupling limit $e/m \ll 1$ at  $T=0 ~ (L=\infty)$,
$\la \cos (2 \pi p_W - \theta ) \ra_f =1$ and 
the boson mass $\mu_1 = 2 e^\gamma m$.  On the other hand, in the 
strong coupling limit $e/m \gg 1$ or in the massless fermion theory,
$\la \cos (2 \pi p_W - \theta ) \ra_f = e^{-\pi/\mu L} \, \cos \theta$.
Eqs.\ (\ref{Schro3}) and (\ref{mass1}) must be solved simultaneously.

The chiral condensate is given by
\beeq
\la \psibar \psi\ra_\theta =  - {2 B(\mu_1L)\over L} 
 ~  \la \cos (2 \pi p_W - \theta ) \ra_f + {e^{2\gamma}\over \pi} \, m ~.
\label{condensate1}
\eneq
The condensate is normalized such that it vanishes in the weak 
coupling limit at $T=0$.  (In the earlier references$^{1-5}$
the chiral condensate has been defined without the last term in 
(\ref{condensate1}).  This reflects ambiguity in defining composite operators.)
In the  massless theory ($m=0$), $\mu_1=\mu$ and
\beeq
\la \psibar\psi \ra_\theta = -{2\over L} B(\mu L) e^{-\pi/\mu L} 
\cos \theta
\label{condensate2}
\eneq. 

\begin{figure}[hbt]
\epsfxsize=9cm  
\centerline{
\epsffile[32 191 476 502]{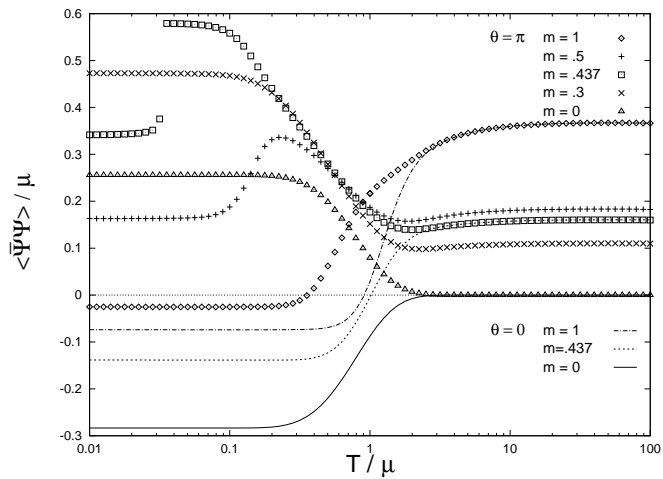}
}
\caption{$T$ dependence of the chiral condensate in the $N=1$ model.
Lines and points are for $\theta=0$ and $\pi$, respectively.
The mass $m$ in the figure is measured in the unit of $\mu$.}
\label{fig:1}
\end{figure}

In fig.\ 1 the condensate $\la \psibar\psi\ra_\theta / \mu$ is plotted
as a function of the temperature $T/\mu = 1/\mu L$.  There is a
crossover transition around $T/\mu \sim 1$. At high temperature
the condensate becomes $\theta$ independent, irrespective of $m/\mu$. 
At low temperature, however, there appears a significant difference
between $\theta=0$ and $\theta=\pi$.

At low $T$ the condensate approaches $-e^\gamma \mu \cos \theta / 2
\pi$ in the massless theory.  It, however, vanishes in the large mass
limit as it approaches a free theory.

At high $T$ the condensate vanishes in the massless theory, whereas it
approaches a non-vanishing value in the massive theory.  Thermal 
excitations yied finite condensates in the massive theory, as there
is no chiral symmetry.  

The $\theta$ dependence of the condensate originates from the
potential.  At low $T/\mu =1/\mu L\ll 1$ it is given by
\beeq
V = \pi^2 \mu^2 L^2 p_W^2 -
 m\mu_1 L^2   \cos(\theta - 2\pi p_W) ~.
\label{potential1}
\eneq
At $\theta=0$ the potential is always minimized at $p_W=0$ and
everything smoothly changes.  As the fermion mass becomes
larger, the condensate $\la \psibar\psi\ra$ increases. 

\begin{figure}[hbt]
\epsfxsize=8cm  
\centerline{
\epsffile[42 291 484 601]{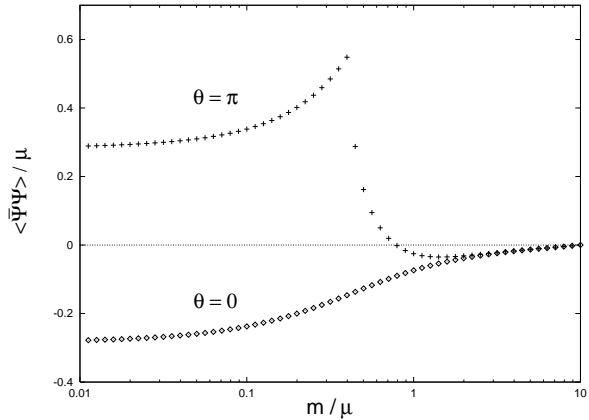}
}
\caption{$m$ dependence of the chiral condensate in the $N=1$ model.
$T/\mu = .03$}
\label{fig:2}
\end{figure}

On the other hand at $\theta=\pi$ there appear two degenetrate minima 
for large $m$.  The problem becomes delicate for a moderate value 
of $m/\mu \sim 1$.
$\mu_1$ is determined by (\ref{mass1}) which involves the vacuum wave
function $f(p_W)$ determined by the potential (\ref{potential1})
itself.   There appear two or three consistent solutions.

For small $m/\mu$, the condensate decreases as $T/\mu$ increases.
The behavior is opposite for large $m/\mu$.  The condensate increases
as $T/\mu$.  For intermediate $m/\mu$, the condensate initially
increases as $T/\mu$, but eventually starts to decrease.   
We recognize profound structure in the behavior of the condensate
for the intermediate coupling $e/m =$O(1) at $\theta=\pi$.

\begin{figure}[hbt]
\epsfxsize=8cm  
\centerline{
\epsffile[42 291 487 602]{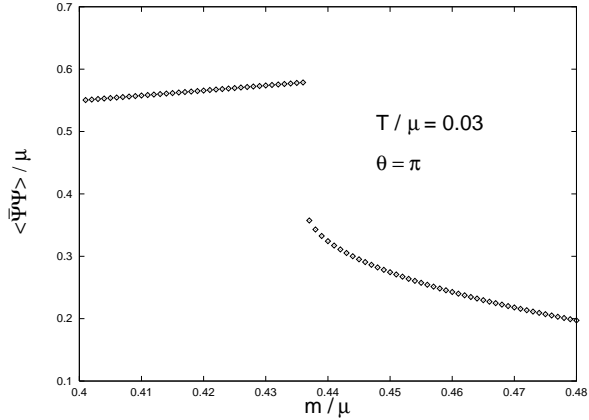}
}
\caption{Discontinuity in the $m$ dependence of the chiral condensate 
in the $N=1$ model at $\theta=\pi$.}
\label{fig:3}
\end{figure}

\section{Anomalous behavior at $\theta=\pi$}

To get more insight into the anomalous behavior of the condensate
near $\theta=\pi$, we have depicted the $m$ dependence of
the condensate at $T/\mu=.03$ in fig.\ 2.  

At $\theta=0$ the absolute value of the chiral condensate at $T=0$
decreases as $m/\mu$ increases, and vanishes at $m/\mu=\infty$.
This result has been previously obtained by Tomachi and Fujita
by the Bogoliubov transformation.\cite{Tomachi}
  Our result agrees
with theirs numerically.  

However, at $\theta=\pi$, there appears a discontinuity in the 
$m/\mu$ dependence.  To magnify this point, more detailed study
is presented in fig.\ 3, in which the condensate is plotted
for $0.4 < m/\mu < .48$ at $T/\mu=.03$.  The discontinuity occurs
at $m/\mu = .437$.

Why and how can such a discontinuity arise in a theory with a 
Hamiltonian having the smooth dependence on  $m$?  To 
understand it, we have  to go back to the equation (\ref{Schro3}).

Suppose that $\mu L=\mu/T$, $m/\mu$, and $\theta$ are given.
The potential $V(p_W)$ in (\ref{Schro3}) is then fixed by
the coefficient $\kappa$ of the $\cos(\theta-2\pi p_W)$ term.
With a given $\kappa_{\rm in}$, the vacuum wave function
$f(p_W)$ is  determined, solving the Schr\"odinger equation.  Now
the boson mass $\mu_1$ (or $\mu_1 L$ numerically) is obtained by
solving (\ref{mass1}) from $f(p_W)$.  The output 
$\kappa_{\rm out}=4\pi mLB(\mu_1 L)$ is determined with
this new $\mu_1$.  
This process defines a mapping $\kappa_{\rm out} = g(\kappa_{\rm in})$:
\beeq
\kappa_{\rm in} \go V(p_W) \go f(p_W) \go \mu_1 L \go \kappa_{\rm out}
~~.
\label{cycle}
\eneq  

We are looking for a solution $\kappa_{\rm out} = \kappa_{\rm in}$,
namely a fixed point of $g(\kappa)$.  Although the function $g(\kappa)$
depends on $m/\mu$ very smoothly,  there takes place 
bifurcation in the fixed point structure as $m/\mu$ varies.

This problem was investigated in ref.\ 5.   It has been found
that for  $T/\mu < 0.12$ and  $\theta=\pi$, there appears two attracting
and one repelling fixed points for $m_c/\mu < m/\mu < m_c'/\mu$. 
Among these three fixed points, the biggest $\kappa$ corresponds 
to the lowest energy density and is chosen.  Hence the lower
critical mass $m_c/\mu$ appears as the location of the discontinuity
in the $m/\mu$ dependence of the various quantities.  In fig.\ 4
we have displayed the $m$ dependence of the $\kappa$ parameter
at $T/\mu=0.03$.

\begin{figure}[hbt]
\epsfxsize=8cm  
\centerline{
\epsffile[58 291 487 602]{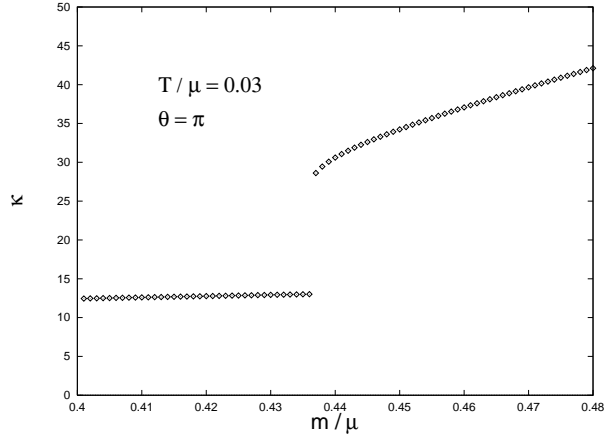}
}
\caption{Discontinuity in the $m$ dependence of the $\kappa$ parameter
in the $N=1$ model at $\theta=\pi$.}
\label{fig:4}
\end{figure}

In general $m_c$ depends on $T$.  At $T=0$ it can be determined 
analytically to be  $m_c/\mu = 0.435$.  At finite $T$ we have found
numerically $m_c/\mu$=0.437 and 0.454 at $T/\mu$=0.03 and 0.07,
repectively.  

This gives us a puzzle.  The Mermin-Wagner theorem ensures that
there is no discontinuity in the $T$ dependence of any physical
quantities.  It implies that if there is a discontinuity in the 
$m$ dependence at $m_c$, $m_c$ must be independent of $T$.
In our approximation we have found that $m_c$ is almost
universal, but has small $T$ dependence.  Indeed, we see a jump in the
$T$ dependence of the chiral condensate for $m/\mu=.437$ in fig.\ 1,
though it may be a smooth transition when the approximation is 
improved.

There are two possible scenarios.  The first possibility is that $m_c$
is universal and $T$ independent in the full theory.  The second possibility
is that the discontinuity in $m$ disappears and is
replaced by a rapid crossover in the full theory.
It is a challenge to know which picture is real.

\section{Generalization and summary}

The generalization to the $N$ flavor case is straightforward.  Extensive
analysis for small fermion masses has been given in
refs.\ 1-4.
For instance, the chiral condensate at low $T$ for
degenerate fermion masses ($m_a=m\ll \mu$) is given by
\beeq
{\mybig 1\over\mybig  \mu} \la \psibar\psi \ra_\theta =
- {\mybig 1\over\mybig  4\pi}
\Big(2 e^\gamma \cos {\mybig \bar\theta\over\mybig  N} \Big)^{{2N\over N+1}} \,
 \Big( {\mybig m\over \mybig \mu} \Big)^{{N-1\over N+1}}
\quad {\rm for} ~ T \ll m^{{N\over N+1}} \mu^{{1\over N+1}}
\eneq 
where $\bar\theta$ is defined in the interval $-\pi\le \bar\theta \le
+\pi$ by
$\bar\theta=\theta-2\pi[(\theta+\pi)/2\pi]$.
Recently Smilga\cite{Smilga} has obtained an exact result for $N=2$  at
$\theta=0$, $T=0$, and $m\ll \mu$, which agrees with our result within
5 \%.  At $T=0$ the cusp singularity appears 
at $\theta=\pi$.  The $N=3$ model with fermion
masses $m_1 < m_2 \ll m_3$ mimics the four-dimensional QCD. 

In this article we have presented an alternative method to explore 
QED in two dimensions.  QED is reduced to a quantum mechanical
system of finite degrees of freedom whose Hamiltonian needs to
be determined by its ground state wave function itself.  
The associated Schr\"odinger equation has been solved numerically
on workstations.  With this algorithm one can determine various
phyisical quantities at any temperature and $\theta$ and with arbitrary
fermion masses.  Our method supplements other numerical
methods such as the lattice gauge theory and light-front quantization
method.$^{8-9}$  More results by these methods
are wanted for comparison.

\section*{Acknowledgments}
This work was supported in part by the U.S.\ Department of Energy
under contracts  DE-AC02-83ER-40105.

\end{document}